\documentclass[
reprint,
superscriptaddress,
amsmath,amssymb, 
aps,
pra,
]{revtex4-2}

\usepackage{dcolumn}
\usepackage{bm}
\usepackage[colorlinks=true,citecolor=red,linkcolor=blue]{hyperref}

\usepackage{diagbox}

\begin{document}

\title{Quantum Information Scrambling and Entanglement: An Elegant Mathematical Connection}

\author{Kapil K. Sharma} \email{iitbkapil@gmail.com, kapil.sharma3@sharda.ac.in} 
\affiliation{School of Basic Sciences and Research, Sharda Univ., Noida, India} 

\author{Rishikant Rajdeepak}
\email{rishikant.rajdeepak@dypiu.ac.in} 
\affiliation{D Y Patil International University, Akurdi, Pune}  

\author{Ashok Kumar}
\email{ashok.kumar6@sharda.ac.in} 
\affiliation{School of Basic Sciences and Research, Sharda Univ., Noida, India} 

\author{Prasanta K. Panigrahi}
\email{pprasanta@iiserkol.ac.in} 
\affiliation{Department of Physical Sciences, Indian Institute of Science Education and Research Kolkata, Mohanpur-741246, West Bengal, India}
\affiliation{Center for Quantum Science and Technology\\
Siksha ’O’ Anusandhan University, Bhubaneswar-751030, Odisha, India}





\begin{abstract}
Studying the behavior of quantum information (QI) scrambling in various quantum systems is an active area of research. Recently, Sharma et al. [K.K. Sharma, V.P Gerdt, Quantum Inf. Process \textbf{20}, 195 (2021)] have shown the mathematical connection between QI scrambling and bipartite entanglement in non-thermal states. In the present work, we elegantly generalize this mathematical connection and study such connections in X-states, non-maximally entangled Bell states, and Werner states.
\end{abstract}


\maketitle


\section{Introduction}
The \cite{Hayden2007,Sekino2008,Lashkari2013,
Landsman2019,Boeing2016,Cencini2010,
Ghys2015,Stockmann1999,Peres1984,
Jalabert2001,Spohn1978,Kitaev2014} phenomenon of QI scrambling can occur in any physical system due to chaotic situations. In classical mechanics, chaos has been studied through the dynamics of trajectories in phase space. If the physical system is very sensitive to initial conditions, the trajectories diverge in space and follow the Lyapunov exponent as $e^{\lambda t}$ \cite{Boeing2016,Cencini2010}. The divergence of trajectories in phase space, following the Lyapunov exponent, leads to what is known as the Butterfly Effect. \cite{Ghys2015}. Studying the butterfly effect in quantum mechanics is challenging due to the absence of the notion of trajectories (excluding de Broglie–Bohm theory). In the quantum domain, quantum chaos is quantified in various ways. One popular approach is to measure the degree of irreversibility by using the mismatch between the forward and backward evolution of the system. In the literature, well-known quantifiers include Loschmidt Echo and irreversible entropy production \cite{Peres1984, Jalabert2001, Spohn1978}. These quantifiers also have experimental manifestations in various physical systems. In the context of the forward-backward evolution approach, out-of-time-ordered correlators (OTOCs) have recently gained much attention for measuring QI scrambling in thermal density matrices \cite{Kitaev2014}. OTOCs were originally discovered by Larkin and Ovchinnikov while studying the quasi-classical method in the theory of superconductivity in 1968 \cite{ot1}. They studied the behavior of the classical pair correlator function, i.e., $C_{c}(t)=\langle [p(t)p(0)]^{\dagger} [p(t)p(0)] \rangle = e^{2\lambda t}$, in a Fermi gas with the Lyapunov exponent $\lambda$. The Lyapunov exponent quantifies the strength of chaos, which is unbounded for classical physical systems, while it is bounded for quantum systems with the limit $\lambda \leq 2\pi k T / \hbar$. Recent trends in quantum chaos focus on the quantum mechanical version of $C_{c}(t)$, represented in terms of quantum operators. Continuing the discussion on OTOCs, it is worth noting that the investigation of different versions of OTOCs is also an active area of research. The impact of OTOCs can disturb quantum correlations in a physical system, but the distinction between QI scrambling and decoherence is not yet clear \cite{qid1}. Extensive work has been conducted on OTOCs in various physical systems across different domains, such as conformal field theories, quantum phase transitions, Luttinger liquids, quantum Ising chains, symmetric Kitaev chains, quadratic fermions, hardcore boson models, and XX spin chains with random fields \cite{cf1, lt1, lt2, qch1, fot1, xot1}. Often, OTOCs in spin chains are studied as a function of the distance between two arbitrary spins, influenced by the actions of local non-commutative operators \cite{scv1}. Furthermore, the Lyapunov exponent as a function of velocity, i.e., $\lambda(v)$, has been studied in classical and semiclassical regimes, and the early-time behavior of QI scrambling has been investigated using the Baker-Campbell-Hausdorff (BCH) formula \cite{bch1, bch2}. Currently, the quantum chaos community is focused on studying OTOCs in thermal density matrices, following the analogy of temperature effects in black holes. However, dynamic studies in various non-thermal quantum states are still lacking.

QI scrambling in any non-thermal physical system can be related to entanglement. In 2021, Sharma et al. demonstrated a mathematical relation between QIS and entanglement in bipartite quantum states. However, this relation is restricted by a few assumptions, such as the density matrix of a given quantum state must have real elements, and the matrix $M$ must satisfy the condition $\text{Tr}(M) = \text{Tr}\left((\sigma_{y} \otimes \sigma_{y}) M\right)$\cite{sh1}. It is very difficult to determine the generalized structure of $M$, but in the present work, we omit these restrictions and generalize the mathematical relation, making it more elegant.

The quantum mechanical version of the OTOC is given by the expectation value of the operator
\begin{equation}
C(t) = [W(t), V]^{\dagger} [W(t), V]. \label{eq:C(t)}
\end{equation}
In quantum mechanics, the expectation value of an operator $O$ is defined as
\[
\langle O \rangle_\rho = \text{Tr}(O \rho),
\]
where $O$ is the operator and $\rho$ represents the density matrix of a quantum state. The expectation value of $C(t)$ is represented in the Heisenberg picture, and it is assumed that the operators $W(t)$ and $V$ are Hermitian as well as unitary, i.e.,
$W(t)=W(t)^{-1}=W(t)^\dagger \mbox{ and } V=V^{-1}=V^\dagger.$ At the initial stage, $t=0$, the operators \(W(0)\) and \(V\) commute, and no QI scrambling takes place. This condition is expressed as
\[
[W(0), V] = 0.
\]
As time progresses, the commutativity between $W(t)$ and $V$ may break, resulting in QI scrambling. Thus, the condition for the existence of QI scrambling can be expressed as
\begin{equation*}
[W(t), V] \neq 0.
\end{equation*}
The unitary time evolution of the operator $W(t)$ under a given Hamiltonian determines the degree of commutativity and, consequently, the extent of information scrambling. The unitary time evolution of the operator $W(t)$ is given by the series
\begin{eqnarray}
 W(t)&=&e^{iHt}W(0)e^{-iHt} \nonumber \\ &=& W(0)+it[H,W(0)]+\frac{t^{2}}{2!}[H,[H,W(0)]] \nonumber \\ && +\frac{it^{3}}{3!}[H,[H,[H,W(0)]]]+\ldots \nonumber
\end{eqnarray}
At $t=0$, the series yields $W(0)$ and no scrambling occurs. For QI scrambling to occur, the following condition must be satisfied:
\[
[H, W(0)] \neq 0.
\]
It is noteworthy that if $H$ and $W(0)$ are bounded operators with $\|H\| \leq \epsilon$ and $\|W(0)\| \leq \epsilon$, then the series is convergent, which facilitates the study of the behavior of QI scrambling.


\section{Linking QI Scrambling with Uhlmann Fidelity and the Bures Metric} 
In this section, we derive the out-of-time-order correlator (OTOC) and elucidate the mathematical connections among QI scrambling, Uhlmann fidelity, and the Bures metric for pure quantum states. We begin by expanding the operator $C(t)$ given by Eq.~(\eqref{eq:C(t)}):
\begin{eqnarray}
    C(t) &=& [W(t), V]^{\dagger} [W(t), V] \nonumber\\
   &=& 2I - \left(W(t) V W(t) V\right)^{\dagger} - W(t) V W(t) V, \nonumber
\end{eqnarray}
where \(I\) is the identity matrix. Taking the expectation value of $C(t)$ with respect to a density matrix \(\rho\) and applying the cyclic property of the trace, the expression simplifies to:
\[
\langle C(t) \rangle_\rho = 2 \left(1 - \Re \left(\text{Tr}(W(t) V W(t) V \rho)\right)\right).
\]
Here, \(\Re\) denotes the real part of the complex quantity. Thus, the OTOC expression obtained is:
\begin{equation}
\langle C(t) \rangle_\rho = 2 \left(1 - \Re (Z(t))\right) \label{eq:<C(t)>}
\end{equation}
where
\[
Z(t) = \text{Tr}(M(t)), \quad M(t) = W(t) V W(t) V \rho. 
\]
To establish the connection between the OTOC and Uhlmann fidelity for pure quantum states, we reconsider Eq.~(\ref{eq:<C(t)>}) for the pure density matrix \(\rho = |\psi \rangle \langle \psi |\). Using the cyclic property of the trace, we rewrite the expression for $\Re(Z(t))$ as:
\begin{eqnarray}
\Re(Z(t)) &=& \Re\left(\langle \psi | W(t) V W(t) V | \psi \rangle\right) \nonumber\\
&=& \Re\left(\langle y(t) | x(t) \rangle\right) \label{eq:Re(Z(t))}
\end{eqnarray}
with
\begin{equation*}
|x(t) \rangle = W(t) V | \psi \rangle \mbox{ and } |y(t) \rangle = V W(t) | \psi \rangle.
\end{equation*}
Here, the state \(|x(t) \rangle\) represents the forward evolution of \(|\psi \rangle\), while \(| y(t) \rangle\) represents the backward evolution of \(|\psi \rangle\) under the actions of \(W(t)\) and \(V\). Since \(W(t)\) and \(V\) are unitary, the state \(|\psi \rangle\) retains its purity throughout the evolution process. The Uhlmann fidelity between two quantum states \(\{|x(t) \rangle, |y(t) \rangle\}\) is defined as:
\begin{equation}
f(t) = |\langle y(t) | x(t) \rangle|^2\,. \label{eq:f(t)}
\end{equation}
This fidelity quantifies the overlap between the forward and backward evolutions of the quantum state described in Eq.~(\ref{eq:<C(t)>}), with \(f(t)\) being a real quantity in the range \(0\leq f(t) \leq 1\). Substituting the expression for $|x\rangle$ and $|y\rangle$ into Eq.~(\ref{eq:f(t)}), we obtain:
\begin{eqnarray}
f(t) &=& |Z(t)|^2 \nonumber\\
 &=& \left( \Re(Z(t)) \right)^2 + \left( \Im (Z(t)) \right)^2. \label{eq:f(t)_2}
\end{eqnarray}
Here, $\Im$ denotes the imaginary part of the complex quantity. 

Next, we consider Eq.~(\eqref{eq:<C(t)>}) in the context of a mixed quantum state described by:
\[
\rho = \sum_{i=1}^n p_i |\psi_i\rangle \langle \psi_i|\,,
\quad n \geq 2\,, \quad p_i \geq 0\,, \quad \sum_{i=1}^n p_i = 1\,,
\]
defined on the Hilbert space \(\mathcal{H}_A\). To facilitate the analysis, we extend \(\rho\) by considering its purification \cite{pu} in the larger Hilbert space \(\mathcal{H}_A \otimes \mathcal{H}_B\), where \(\mathcal{H}_A\) represents the original Hilbert space and \(\mathcal{H}_B\) denotes an auxiliary Hilbert space used for the purification process. The purification is given by:
\[
|\Psi\rangle = \sum_{i=1}^n \sqrt{p_i} |\psi_i\rangle |\phi_i\rangle\,, \quad \text{with} \quad \text{Ptr}_B \left(|\Psi\rangle \langle \Psi|\right) = \rho,
\]
where \(\text{Ptr}_B\) denotes the partial trace over the subsystem \(B\). Thus, the quantity $Z(t)$ in the Eq.~(\eqref{eq:<C(t)>}) can be rewritten as:
\[
Z(t) = \langle \Psi | \left( W(t) V W(t) V \otimes I_B \right) |\Psi \rangle\,,
\]
where \(I_B\) is the identity matrix acting on \(\mathcal{H}_B\). Using analogous reasoning as in Eq.~(\ref{eq:Re(Z(t))}), we can use the pure states:
\[
W(t) V \otimes I_B |\Psi\rangle \quad \text{and} \quad V W(t) \otimes I_B |\Psi\rangle\,.
\]
Hence, Eqs.~\eqref{eq:<C(t)>} and \eqref{eq:f(t)_2} hold true for both pure and mixed states. We now use these equations to derive the following:
\begin{equation}\label{eq:R(Z(t))_2}
\Re(Z(t)) = \sqrt{f(t) - \left( \Im(Z(t)) \right)^2}
\end{equation}
and
\begin{equation}
\langle C(t) \rangle_\rho = 2 \left( 1 - \sqrt{f(t) - \left( \Im(Z(t)) \right)^2} \right).
\label{eq:<C(t)>&f(t)}
\end{equation}
Equation~\eqref{eq:<C(t)>&f(t)} establishes the relationship between Uhlmann fidelity and QI scrambling. When the imaginary part of \(Z(t)\) is zero, \(\langle C(t) \rangle_\rho\) simplifies to:
\begin{equation}
\langle C(t) \rangle_\rho = 2 \left( 1 - \sqrt{f(t)} \right) \,. \label{eq:<C(t)>&f(t)_2}
\end{equation}
Equations~\eqref{eq:R(Z(t))_2} and~\eqref{eq:<C(t)>&f(t)_2} show that \(\sqrt{f(t)}\) is the exact upper bound for \(\Re(Z(t))\), while \(2(1 - \sqrt{f(t)})\) is the exact lower bound for the QI scrambling \(\langle C(t) \rangle_\rho\). These bounds are attained when \(Z(t)\) is real. 

We now explore the connection between QI scrambling and the Bures metric. According to~\cite{bl2}, the relationship between the Bures metric \(D\) and the Uhlmann fidelity is given by:
\begin{equation}
D(t) = \sqrt{2 \left( 1 - \sqrt{f(t)} \right)} \label{eq:D(t)&f(t)}
\end{equation}
Substituting \(f(t)\) from Eq.~(\ref{eq:D(t)&f(t)}) into Eq.~(\ref{eq:<C(t)>&f(t)}), we obtain:
\begin{equation}
\langle C(t) \rangle_\rho = 2 \left( 1 - \sqrt{\left( 1 - \frac{D(t)^2}{2} \right)^2 - \left( \Im(Z(t)) \right)^2} \right) \label{eq:<C(t)>&D(t)}
\end{equation}
When \(Z(t)\) is real, Eq.~(\ref{eq:<C(t)>&D(t)}) simplifies to:
\[
\langle C(t) \rangle_\rho = D(t)^2\,.
\]
This final equation describes the relationship between QI scrambling and the Bures metric. The Bures metric provides an alternative measure of the closeness of two quantum states and is related to the Uhlmann fidelity. We find that, for real \(Z(t)\), QI scrambling is a square function of the Bures metric. 

We conclude this section by highlighting the properties of QI scrambling derived from Eq.~(\ref{eq:<C(t)>&f(t)}):
\begin{enumerate}
    \item \textit{Positivity:} \(\langle C(t) \rangle_\rho \geq 0\)
    \item \textit{Bounded Limits:} \(0 \leq \langle C(t) \rangle_\rho \leq 2\)
    \item \textit{Unitary Invariance:} \(U \langle C(t) \rangle_\rho U^{\dagger} = \langle C(t) \rangle_\rho\)
\end{enumerate}
Moreover, it is well known that Uhlmann fidelity possesses the symmetry:
\[
f(t) = |\langle y(t) | x(t) \rangle|^2 = |\langle x(t) | y(t) \rangle|^2\,.
\]
Therefore, we conclude that QI scrambling \(\langle C(t) \rangle_\rho\) respects this symmetry under the exchange of forward and backward evolution. In this paper, we emphasize Uhlmann fidelity due to its well-established properties and widespread use in the literature.


\section{Linking QI scrambling and entanglement}
In this section, we establish the mathematical connection between QI scrambling and bipartite concurrence. The concurrence, a measure of entanglement, is defined as:
\begin{equation*}
C_{r}(|\psi\rangle) = |\langle\psi|\sigma_{y} \otimes \sigma_{y}|\psi^{\star}\rangle|,
\end{equation*}
or equivalently,
\begin{equation*}
C_{r}(|\psi\rangle) = \left| \text{Tr}\left((\sigma_{y} \otimes \sigma_{y}) (|\psi^{\star}\rangle\langle\psi|)\right) \right|, 
\end{equation*}
where $(\sigma_{y} \otimes \sigma_{y})$ is the spin-flip operator that acts on both qubits, and $|\psi^{\star}\rangle$ is the complex conjugate of the state $|\psi\rangle$. In the case where $|\psi^{\star}\rangle = |\psi\rangle$, , implying that the density matrix $\rho$ is real $(\rho^\star=\rho)$,  we can express the concurrence as:
\begin{equation*}
C_{r}(\rho) = |\text{Tr}\left((\sigma_{y} \otimes \sigma_{y}) \rho\right)|.
\end{equation*}
We can extend this definition to the concurrence in the state $M(t)$ after evolution, as given by:
\begin{eqnarray}
C_{r}(M(t)) &=& |\text{Tr}\left((\sigma_{y} \otimes \sigma_{y}) M(t)\right)| \nonumber
\end{eqnarray}
Although, $M(t)$ is generally non-Hermitian, it remains Hermitian when the Hamiltonian $H$ is of the form $H=cI$. We explore the cases where \[|\text{Tr}\left((\sigma_{y} \otimes \sigma_{y}) M(t)\right)|\leq |\text{Tr}(M(t))|,\] leading to the expression:
\begin{eqnarray}
    C_r(M(t)) &=& k  |\text{Tr}(M(t))|, \mbox{ with }0\leq k\leq 1. \label{eq:k} \\
    &=& k \sqrt{f(t)} \label{eq:C_r&f(t)}.
\end{eqnarray}
This equation shows a non-linear relationship between concurrence in the state $M(t)$ and the function $f(t)$, scaled by the constant $k$.

The complex matrix $M(t)$, satisfying the trace invariance under  $(\sigma_{y} \otimes \sigma_{y})$, has the form:
\begin{equation}
M(t) = \left(
\begin{array}{cccc}
 a & b & c & -a \\
 d & e & e & f \\
 g & h & h & i \\
 j & k & l & -j \\
\end{array}
\right) \label{xt}
\end{equation}
The exact structure of $M(t)$ is influenced by the Hamiltonian $H$, the scrambling operators $W(t)$ and $V$, and the density matrix $\rho$. Determining the precise form of these components is a challenging task.

By obtaining $\sqrt{f(t)}$ from Eq.~\eqref{eq:C_r&f(t)} and plugging it into Eq.~\eqref{eq:D(t)&f(t)}, we can establish the direct connection between concurrence and Bures metric as:
\[
C_{r}(M(t)) = k \left(1 - \frac{D(t)^{2}}{2}\right)\,.
\]
Similarly, by substituting $f(t)$ into Eq.~\eqref{eq:<C(t)>&f(t)}, we establish the direct connection between QI scrambling and concurrence as:
\[
\langle C(t)\rangle_\rho = 2\left(1 - \sqrt{\left(\frac{C_{r}(M(t))}{k}\right)^{2} - \left(\Im(Z(t))\right)^{2}}\right)\,. \label{cm1}
\]
Rearranging the above equation gives:
\begin{equation}
C_{r}(M(t)) = k \sqrt{\left(1 - \frac{\langle C(t)\rangle_\rho}{2}\right)^{2} + \left(\Im(Z(t))\right)^{2}}\,. \label{eq:C_r&<C(t)>}
\end{equation}
The above equation links QI scrambling with concurrence. For $\Im(Z(t))=0$, it simplifies to
\[C_r(M(t)) = k \left(1 - \frac{\langle C(t)\rangle_\rho}{2}\right).\]
These mathematical relations are essential for understanding the influence of QI scrambling on two-qubit concurrence during evolution. These expressions are generalized for any density matrix with real elements, provided the condition in Eq.~\eqref{eq:k} is met.


\section{QI Scrambling and Entanglement in different quantum states}
In this section, we explore the mathematical relationship between QI scrambling and entanglement, focusing on the scenario involving a diagonal Hamiltonian. Specifically, we consider the Ising Hamiltonian:
\begin{equation*}
H = -J_z \sum_{i,j=1}^{2} \sigma^z_i \sigma^z_j.
\end{equation*}
We examine the behavior of operators $W(0)$ and $V$, which act on the first qubit and are drawn from the set $\{\sigma_i, \sigma_j\}$. Within the composite Hilbert space $\mathcal{H}_1 \otimes \mathcal{H}_2$, these operators assume the form:
\begin{equation}
\{W(0) \otimes I, V \otimes I\} \subseteq \{\sigma_i \otimes I, \sigma_j \otimes I\}. \label{eq:W&V_subset}
\end{equation}
The time evolution of the operator $W(0)$ is governed by $W(t) = U(t)^{\dagger} W(0) U(t)$, where $U(t) = e^{-iHt}$. To establish the relationship between QI scrambling and entanglement, we begin by preparing two qubits in symmetric X states, non-maximally entangled Bell states, and Werner states. Subsequent subsections analyze the behaviour of the parameter $k$, which serves as the key to linking QI scrambling and entanglement. It is important to note that this study is conducted within the framework of a diagonal Hamiltonian, specifically the Ising Hamiltonian.


\subsection{ QI Scrambling and Entanglement in Symmetric X States}
\label{subsec:X-state}
In this subsection, we develop the mathematical relationship between QI scrambling and entanglement in symmetric X states. These states are represented by the density matrix:
\begin{equation}
\rho^{x}=\begin{pmatrix}a & 0 & 0 & w\\
0 & b & z & 0\\
0 & z & c & 0\\
w & 0 & 0 & d
\end{pmatrix}, \label{eq:densitymatrix}
\end{equation}
which satisfies the normalization condition:
\begin{equation}
(a+b+c+d)=1. \label{eq:trace}
\end{equation}
We compute the parameter $k$ for different combinations of $\{W(0),V\}$, as shown in  Table~\ref{table:valueofk_1}. The parameter $k$ takes the value $(k=\tilde{l})$ for the following combinations:
\begin{equation}
\{W(0),V\}=
\begin{cases}
(\sigma_{x}\otimes I,\sigma_{x}\otimes I),\\
(\sigma_{x}\otimes I,\sigma_{y}\otimes I),\\
(\sigma_{y}\otimes I,\sigma_{x}\otimes I),\\
(\sigma_{y}\otimes I,\sigma_{y}\otimes I) .
\end{cases}\label{eq:W(0)&V_combination_1}
\end{equation}  
The parameter $\tilde{l}$ is expressed as:
\begin{equation}
\tilde{l}=\frac{2\sqrt{w^{2}+z^{2}-2wz\cos(\phi)}}{\sqrt{(b+c)^{2}+(a+d)^{2}+2(b+c)(a+d)\cos(\phi)}}, \label{eq:expressionof_l} 
\end{equation}
with $(\phi=8J_{z}t)$. The value of $k=\tilde{l}$ is substituted into  
Eq.~(\ref{eq:C_r&<C(t)>}) to obtain the relation between QI scrambling and entanglement. The simplified version of this relation, under the condition ${\Im}(Z)=0$, is given by:
\begin{equation}
C_{r}(M(t))=\tilde{l}.\left(1-\left(\frac{\langle C(t)\rangle_\rho}{2} \right)\right). \label{eq:C_r(M)_expressionof_l}
\end{equation}

For the following combinations:
\begin{equation}
\{W(0),V\}=
\begin{cases}
(\sigma_{z}\otimes I,\sigma_{x}\otimes I),\\
 (\sigma_{z}\otimes I,\sigma_{y}\otimes I),\\
 (\sigma_{z}\otimes I,\sigma_{z}\otimes I),
\end{cases} \label{pc2}
\end{equation}
the parameter $k$ takes the value $k=\tilde{\tilde{l}}=2(w-z)$. Using the same procedure as above, the relationship between QI scrambling and entanglement is obtained as:
\begin{equation}
C_{r}(M(t))=\tilde{\tilde{l}}.\left(1-\left(\frac{\langle C(t)\rangle_\rho}{2} \right)\right). \label{eq:C_r(M)_expressionof_second_l}
\end{equation}
It is worth noting that $\tilde{l}$ coincides with $\tilde{\tilde{l}}$, i.e., $\tilde{l}=\tilde{\tilde{l}}=2(w-z)$, when $t=0$, or equivalently $\phi=0$. One can simulate the above expressions to investigate the behavior of entanglement versus QI scrambling for \( t > 0 \), with the condition \( 0 \leq \tilde{l} \leq 1 \). This condition imposes constraints on the parameters involved in the equation, including the density matrix parameters and \( \phi \).

To explore this further, we consider Eq.~(\ref{eq:f(t)_2}) and analytically derive the constraints on the density matrix parameters. We aim to identify the structure of \( \rho^{x} \) that satisfies the condition \( 0 \leq \tilde{l} \leq 1 \). For an arbitrary solution, we impose the restriction \( w = a \) and \( z = d \). Using the trace condition in Eq.~(\ref{eq:trace}), we obtain \( b + c = 1 - (a + d) \). Substituting the values of \( w \), \( z \), and \( b + c \) into Eq.~(\ref{eq:expressionof_l}), the equation becomes:
\begin{equation}
\begin{split}
\tilde{l}= \frac{2\sqrt{a^{2}+d^{2}-2ad\cos(\phi)}}{\sqrt{(1-(a+d))^{2}+(a+d)^{2}+2(1-(a+d))(a+d)\cos(\phi)}} \label{eq:restricted_expressionof_l}
\end{split}
\end{equation}   

The factor \( \cos(\phi) \) dictates the time evolution. We fix the time evolution window by assuming \( 0 \leq \cos(\phi) \leq 1 \). We now determine the values of the parameters \( a \), \( d \), and \( \cos(\phi) \) for which \( \tilde{l} \) reaches its lower bound of 0 and upper bound of 1. This analysis will be divided into two sub-cases.

\subsubsection*{Case 1:}

First, we consider the case where \( \cos(\phi) = 0 \) and substitute this value into Eq.~(\ref{eq:restricted_expressionof_l}), under the condition \( \tilde{l} = 0 \), along with the trace condition given in Eq.~(\ref{eq:trace}). The equation simplifies to:
\[
\tilde{l} = a^2 + d^2 = 0.
\]
This directly implies \( a = d = 0 \).

Next, we consider the case where \( \cos(\phi) = 0 \), and substitute this value into Eq.~(\ref{eq:restricted_expressionof_l}), under the condition \( \tilde{l} = 1 \). The equation transforms into:
\[
(a - d)^2 + (a + d) = \frac{1}{2}.
\]
A possible solution to this equation is \( a = \frac{1}{4} \) and \( d = \frac{1}{4} \). By substituting these parameter values into Eq.~(\ref{eq:trace}), we obtain \( b = \frac{1}{4} \) and \( c = \frac{1}{4} \). After substitution of these parameters into Eq.~(\ref{eq:densitymatrix}), the density matrix takes the following form:
\[
\rho^{x} =
\begin{pmatrix}
\frac{1}{4} & 0 & 0 & \frac{1}{4} \\
0 & \frac{1}{4} & \frac{1}{4} & 0 \\
0 & \frac{1}{4} & \frac{1}{4} & 0 \\
\frac{1}{4} & 0 & 0 & \frac{1}{4}
\end{pmatrix}. \label{eq:density_case1}
\]
This density matrix represents a special case of Bell diagonal states. Thus, we have analytically demonstrated that the upper bound, \( \tilde{l} = 1 \), exists for the density matrix given in Eq.~(\ref{eq:densitymatrix}) under this special case.

\begin{table}{}
\centering
\begin{tabular}{c|c|c|c}
\diagbox{W(0)}{V}  & $(\sigma_{x}\otimes I)$ & $(\sigma_{y}\otimes I)$ & $\sigma_{z}$\tabularnewline
\hline 
\hline 
$(\sigma_{x}\otimes I)$ & $\tilde{l}$ & $\tilde{l}$ & $\tilde{\tilde{l}}$\tabularnewline
\hline 
$(\sigma_{y}\otimes I)$ & $\tilde{l}$ & $\tilde{l}$ & $\tilde{\tilde{l}}$\tabularnewline
\hline 
$(\sigma_{z}\otimes I)$ & $\tilde{l}$ & $\tilde{l}$ & $\tilde{\tilde{l}}$\tabularnewline
\hline 
\end{tabular} 
\caption{\label{table:valueofk_1}Values of the parameter $k$ for different operator combinations }
\end{table} 
\subsubsection*{Case 2:}

In this case, we also investigate the constraints on the parameters of the density matrix \( \rho^{x} \), for which the condition \( 0 \leq \tilde{l} \leq 1 \) is satisfied. In Eq.~(\ref{eq:restricted_expressionof_l}), we substitute \( \cos(\phi) = 1 \) with \( \tilde{l} = 0 \), and applying the trace condition, i.e., \( a + b + c + d = 1 \), the equation reads as:
\[
\tilde{l} = 2(a - d) = 0.
\]
This leads to \( a = d \). After substituting \( a = d \) into the trace condition, we obtain \( b + c = 1 - 2a \), with the constraint \( a \leq \frac{1}{2} \). These constraints lead to the lower bound \( \tilde{l} = 0 \).

Secondly, we consider \( \cos(\phi) = 1 \) with \( \tilde{l} = 1 \), and the equation reads as:
\[
\tilde{l} = 2(a - d) = 1.
\]
This leads to \( a - d = \frac{1}{2} \). After substituting \( a = \frac{1}{2} + d \) into the trace condition, we obtain \( b + c = \frac{1}{2}(1 - 4a) \) with the constraint \( 4a \leq 1 \). By imposing these constraints, the entire spectrum of density matrices \( \rho^{x} \) can be obtained for the upper bound \( \tilde{l} = 1 \), validating the relation between QI scrambling and entanglement.

Here, we recall the linear relationship established between QI scrambling and entanglement, obtained in Eqs.~(\ref{eq:C_r(M)_expressionof_l}) and (\ref{eq:C_r(M)_expressionof_second_l}) with the parameters \( \tilde{l} \) and \( \tilde{\tilde{l}} \), respectively. Both equations represent straight lines, where the values of the parameters \( \tilde{l} \) and \( \tilde{\tilde{l}} \) represent the tangent and the intercept at the $Y$-axis. As QI scrambling increases within the limits \( 0 \leq \langle C(t)\rangle \leq 2 \), entanglement decreases linearly. When \( \langle C(t)\rangle = 0 \), the maximum values achieved by the entanglement are governed by the cut at the $Y$-axis, achieved by the values \( \tilde{l} = 1 \) and \( \tilde{\tilde{l}} = 1 \). Hence, entanglement lies within the limits \( 0 \leq C_{r}(M(t)) \leq 1 \).


\begin{table}[t]
\centering
\begin{tabular}{c|c|c|c}
\diagbox{W(0)}{V}  & $(\sigma_{x}\otimes I)$ & $(\sigma_{y}\otimes I)$ & $\sigma_{z}$\\
\hline \hline
$(\sigma_{x}\otimes I)$ & $\tilde{k}$ & $\tilde{k}$ & $\tilde{k}$\\
\hline 
$(\sigma_{y}\otimes I)$ & $\tilde{k}$ & $\tilde{k}$ & $\tilde{k}$\\
\hline 
$(\sigma_{z}\otimes I)$ & $\tilde{k}$ & $\tilde{k}$ & $\tilde{k}$\\
\hline 
\end{tabular}
\caption{\label{table:non-maximally}Values of the parameter $k$ for different operator combinations } 
\end{table}

\subsection{QI Scrambling and Entanglement in Non-Maximally Entangled Bell States}\label{subsec:non-max}

In this sub-section, we obtain the mathematical expression relating QI scrambling and entanglement in non-maximally entangled Bell states. These quantum states are given as:

\begin{eqnarray}
|\phi_{\alpha}^{+}\rangle &=& \frac{1}{\sqrt{1+|\alpha|^{2}}}(|00\rangle+\alpha|11\rangle) \\
|\phi_{\alpha}^{-}\rangle &=& \frac{1}{\sqrt{1+|\alpha|^{2}}}(\alpha^{\star}|00\rangle-|11\rangle) \\
|\psi_{\alpha}^{+}\rangle &=& \frac{1}{\sqrt{1+|\alpha|^{2}}}(|01\rangle+\alpha|10\rangle) \\
|\psi_{\alpha}^{-}\rangle &=& \frac{1}{\sqrt{1+|\alpha|^{2}}}(\alpha^{\star}|01\rangle-|10\rangle)
\end{eqnarray}

Here, \( (\alpha, \beta) \) are generally complex numbers. When \( \alpha = \beta = 1 \), the states reduce to Bell states, which possess maximal entanglement, i.e., 1 ebit.

We prepare the two qubits initially in non-maximally entangled Bell states and calculate the value of the parameter \( k \) from Eq.~(\ref{eq:k}). The values are shown in Table~\ref{table:non-maximally} for different combinations of \( \{W(0), V\} \), see Eq.~\ref{eq:W&V_subset}. We have found that for all operator combinations, the parameter \( k \) achieves the value \( k = \tilde{k} \), given as:

\begin{eqnarray}
\tilde{k} = \frac{2\alpha}{(1 + \alpha^{2})}
\end{eqnarray}

By substituting the value of the parameter \( \tilde{k} \) in Eq.~\ref{eq:C_r&<C(t)>}, we can obtain the relation between QI scrambling and entanglement. A simplified version of this relation, under the condition \( \Im(Z) = 0 \), reads as:

\begin{equation}
C_{r}(M(t)) = \tilde{k} \left(1 - \frac{\langle C(t) \rangle_\rho}{2} \right)
\end{equation}

We have found an interesting result: for all non-maximally entangled Bell states, the parameter \( \tilde{k} \) is independent of time \( t \) under the diagonal Hamiltonian \( H \). There is a linear relationship between entanglement and QI scrambling: as QI scrambling increases, entanglement decreases linearly.

The parameter \( \tilde{k} \) follows the bounds \( 0 \leq \tilde{k} \leq 1 \) with \( 0 \leq \alpha \leq 1 \). For \( \alpha = 0 \), all non-maximally entangled states are separable and carry zero entanglement, while for \( \alpha = 1 \), all states correspond to Bell states and achieve the highest entanglement, i.e., 1 ebit.

\subsection{Werner State:}\label{subsec:werner}
In this section, we establish the mathematical connection between QI scrambling and entanglement in Werner states. The two-qubit Werner state can be written as:
\begin{equation}
\rho_{w} = \gamma |\psi^{-}\rangle \langle \psi^{-}| + (1-\gamma)\frac{I}{4}, \label{wer1}
\end{equation}
where $|\psi^{-}\rangle$ is the singlet state and $I$ is the $4 \times 4$ identity matrix. If the state does not pass through OTOC, then the entanglement carried by the state is \(C(\gamma) = \frac{(3\gamma-1)}{2}\) for \(\gamma > \frac{1}{3}\). However, since in Eq.~\ref{eq:k}, the state evolves under the OTOC influenced by the Hamiltonian $H$, the entanglement in the state will change.

\begin{table}[t]
\begin{tabular}{c|c|c|c}
\diagbox{$W(0)$}{$V$}  & $(\sigma_{x}\otimes I)$ & $(\sigma_{y}\otimes I)$ & $(\sigma_{z}\otimes I)$\tabularnewline
\hline 
\hline 
$(\sigma_{x}\otimes I)$ & $\tilde{m}$ & $\tilde{m}$ & $\tilde{\tilde{m}}$\tabularnewline
\hline 
$(\sigma_{y}\otimes I)$ & $\tilde{m}$ & $\tilde{m}$ & $\tilde{\tilde{m}}$\tabularnewline
\hline 
$(\sigma_{z}\otimes I)$ & $\tilde{m}$ & $\tilde{m}$ & $\tilde{\tilde{m}}$\tabularnewline
\hline 
\end{tabular} 
\caption{\label{table:wernerstate}Values of the parameter $k$ for different operator combinations }
\end{table}

The values of the parameter \(k\) for different combinations of $\{W(0),V\}$ (see Eq.~\ref{eq:W&V_subset}) are shown in Table~\ref{table:wernerstate}. The parameter \(k\) achieves the value \(k = \tilde{m}\) for the combinations given in Eq.~(\ref{eq:W(0)&V_combination_1}), and for the combinations given in Eq.~(\ref{pc2}), the value is \(k = \tilde{\tilde{m}}\). The values of \(\tilde{m}\) and \(\tilde{\tilde{m}}\) are given in the following equations, respectively:
\begin{eqnarray}
\tilde{m} &=& \frac{\gamma}{\sqrt{\cos^{2}(\phi) + \gamma\sin^{2}(\phi)}} \\
\tilde{\tilde{m}} &=& \gamma
\end{eqnarray}
with \(\phi = 4J_{z}t\). For \(t = 0\), both parameters are equal, i.e., \(\tilde{m} = \tilde{\tilde{m}}\). Substituting these values in Eq.~(\ref{eq:C_r&<C(t)>}), the relation between QI scrambling and entanglement can be obtained. The simplified versions with \(\Im(Z) = 0\) are shown below for the conditions \(k = \tilde{m}\) and \(k = \tilde{\tilde{m}}\), respectively:
\begin{eqnarray}
C_{r}(M(t)) = \tilde{m} \left(1 - \frac{\langle C(t)\rangle_\rho}{2} \right) \label{up1} \\
C_{r}(M(t)) = \tilde{\tilde{m}} \left(1 - \frac{\langle C(t)\rangle_\rho}{2} \right) \label{up2}
\end{eqnarray}

We have found that the lower bounds \((\tilde{m} = 0, \tilde{\tilde{m}} = 0)\) are achieved with \(\gamma = 0\), so the corresponding structure of Eq.~(\ref{wer1}) is:
\[
\rho^{w} = (1 - \gamma)\frac{I}{4}.
\]
On the other hand, the upper bounds \((\tilde{m} = 1, \tilde{\tilde{m}} = 1)\) are achieved by solving Eqs.~(\ref{up1}) and (\ref{up2}), which gives \(\gamma = 1\). Thus, the corresponding structure of \(\rho^{w}\) is:
\[
\rho^{w} = \gamma |\psi^{-}\rangle \langle \psi^{-}|.
\]

\section{Conclusion}
In this work, we established the mathematical connections between QI scrambling and entanglement in bipartite quantum states using pure state Uhlmann fidelity. Additionally, we explored the connection between QI scrambling and the Bures metric, further linking it to Uhlmann fidelity. The time evolution of entanglement under the diagonal Ising Hamiltonian was analyzed alongside QI scrambling in various quantum states, including X states, non-maximally entangled Bell states, and Werner states. 

Our investigation revealed that in all these states, entanglement and QI scrambling follow a linear relationship. Interestingly, we found that in non-maximally entangled Bell states, the evolution of entanglement remains unaffected by the time parameter. These mathematical connections between QI scrambling and entanglement were validated across different quantum states. 

This work opens up a broader future scope for exploring similar mathematical relations in other quantum states. Such investigations could enhance our understanding of the direct influence of quantum chaos on entanglement, making this research relevant to the quantum information and computation community.

\bibliography{Reference}

\begin{thebibliography}{}
 \bibitem{Hayden2007} P. Hayden and J. Preskill, Black holes as mirrors: Quantum information in random subsystems, JHEP 0709, 120 (2007).
 \bibitem{Sekino2008} Y. Sekino and L. Susskind, Fast Scramblers, JHEP 0810, 065  (2008).
 \bibitem{Lashkari2013} N. Lashkari, D. Stanford, M. Hastings, T. Osborne, and P. Hayden, Towards the fast scrambling conjecture,  JHEP 1304, 022 (2013).
 \bibitem{Landsman2019}
 K. A. Landsman, C. Figgatt, T. Schuster, N. M. Linke, B. Yoshida, N. Y. Yao and C. Monroe, Verified quantum information scrambling, Nature volume 567, 61 (2019).
 \bibitem{Boeing2016}
 G. Boeing, Visual Analysis of Nonlinear Dynamical Systems: Chaos, Fractals, Self-Similarity and the Limits of Prediction, Systems. 4, 37 (2016).
 \bibitem{Cencini2010}
 M. Cencini et al. Chaos: From Simple models to complex systems, World Scientific (2010).
 \bibitem{Ghys2015}
 É. Ghys, ``The Butterfly Effect,'' in \textit{The Proceedings of the 12th International Congress on Mathematical Education}, edited by S. J. Cho (Springer International Publishing, Cham, 2015), pp. 19--39.
 \bibitem{Stockmann1999}
 H.J Stokmann, Quantum Chaos: An Introduction, Cambridge University Press, (1999). 
 \bibitem{Peres1984}
 A. Peres, Stability of quantum motion in chaotic and regular systems, Phys. Rev. A, 30, 1610 (1984).
 \bibitem{Jalabert2001}
 R. A. Jalabert and H. M. Pastawski, Environment-independent decoherence rate in classically chaotic systems”. Phys. Rev. Lett. 86, 2490 (2001).
 \bibitem{Spohn1978}
 H. Spohn, Entropy production for quantum dynamical semigroups, 	J. Math. Phys. 19, 1227 (1978).
 \bibitem{Kitaev2014}
 A. Kitaev. Hidden Correlations in the Hawking Radiation and Thermal Noise. Talk at the 2015 Breakthrough Prize Fundamental Physics Symposium. Nov. 10, 2014 (2014). 
 \bibitem{ot1}
 A. Larkin and Yu. N. Ovchinnikov, Quasiclassical Method in the Theory of Superconductivity, Sov. Phys. JhETP 28, 6, 
 1200 (1969).
 \bibitem{qid1}
 B. Yoshida and N. Y. Yao, Disentangling Scrambling and Decoherence via Quantum Teleportation, Phys. Rev. X, 9, 011006 (2019).
 \bibitem{cf1}
 D. A. Roberts and D. Stanford, Diagnosing chaos using
 Four-Point functions in Two-Dimensional conformal field
 theory,  Phys. Rev. Lett. 115, 131603 (2015).
 \bibitem{lt1}
 B. Dóra and R. Moessner, Out-of-Time-Ordered Den-sity Correlators in Luttinger Liquids, Phys. Rev. Lett, 119, 026802 (2017).
 \bibitem{lt2}
 C. J. Lin and O. I. Motrunich, Out-of-time-ordered cor-relators in short-range and long-range hard-core boson models and in the Luttinger-liquid model, Phys. Rev. B 98, 134305 (2018).
  \bibitem{qch1}
 C. J. Lin and O. I. Motrunich, Out-of-time-ordered correlators in a quantum Ising chain, Phys. Rev. B 97, 144304 (2018).
 \bibitem{fot1}
 P. Kos and T. Prosen, Time-dependent correlation functions in open quadratic fermionic systems, Time-dependent correlation functions in open quadratic fermionic systems, J. Stat. Mech. 123103 (2017).
 \bibitem{xot1}
 J. Riddell and E. S. Sørensen,
 Out-of-time ordered correlators and entanglement growth in the random-field XX spin chain, Phys. Rev. B 99, 054205 (2019).
 \bibitem{scv1}
 B. Swingle, Unscrambling the physics of out-of-time-order correlators, Nature Physics 14, 988 (2018).
 \bibitem{bch1}
 R. Achilles and A. Bonfiglioli, The early proofs of the theorem of Campbell, Baker, Hausdorff, and Dynkin,
 Arch. Hist. Exact Sci., 66 295 (2012).
 \bibitem{sh1}
  Sharma, K.K., Gerdt, V.P. Quantum information scrambling and entanglement in bipartite quantum states. Quantum Inf. Process 20, 195 (2021).
 \bibitem{pu}
 Bennett, C.H. et al. Purification of Noisy Entanglement and Faithful Teleportation via Noisy Channels, Phys. Rev. Lett. 76, (1996).

 \bibitem{bch2}
 A. Bonfiglioli and R. Fulci, Topics in Noncommutative Algebra: The Theorem of Campbell, Baker, Hausdorff and Dynkin, Lecture Notes in Mathematics, Springer, London (2012).
 \bibitem{con1}
 S. Hill and W. K. Wootters, Entanglement of a Pair of Quantum Bits, Phys. Rev. Lett. 78, 5022 (1997).
 \bibitem{con2}
 W. K. Wootters, Entanglement of Formation of an Arbitrary State of two Qubits, Phys. Rev. Lett. 80, 2245 (1998).
 \bibitem{con3}
 R. Horodecki, P. Horodecki, M. Horodecki and K. Horodecki, Quantum entanglement, Rev. Mod. Phys. 81, 865 (2009).
 \bibitem{uh1}
 A. Uhlmann, Transition Probability (Fidelity) and Its Relatives, Foundations of Physics, 41, 288 (2011).
 \bibitem{bm1}
 D. Bures, An extension of Kakutani's theorem on infinite product measures to the tensor product of semifinite $w\sp{\ast}$-algebras, Trans. Amer. Math. Soc. 135, 199 (1969).
 \bibitem{bl1}
 D. Sych and G. Leuchs, A complete basis of generalized Bell states, 	New J. Phys. 11, 013006 (2009).
 \bibitem{bl2}
 W. K. Wootters, Statistical distance and Hilbert space,  Phy. Rev. 
 D., 23, 357 (1981). 
 \end{thebibliography}

\end{document}